\documentclass[AMA,Times2COL]{WileyNJDv5}

\usepackage{comment}
\usepackage{moresize}
\usepackage{listings}
\usepackage{xcolor}
\usepackage{array}
\usepackage{csquotes}
\usepackage{amsmath,soul}
\usepackage{soulpos}
\usepackage{multirow}
\usepackage{float}
\usepackage{graphicx}
\usepackage{ifthen}

\definecolor{darkgreen}{rgb}{0.0,0.4,0.0} 
\definecolor{darkred}{rgb}{0.6,0.1,0.1}
\definecolor{lightgray}{gray}{.98}
\definecolor{medgray}{gray}{.70}
\definecolor{darkgray}{gray}{.40}
\definecolor{lightviolet}{rgb}{0.7,0,0.7} 
\definecolor{lightlightviolet}{rgb}{1.0,0.7,1.0} 
\definecolor{darkviolet}{rgb}{0.5,0.1,0.5}
\definecolor{darkredviolet}{rgb}{0.6,0.1,0.4}
\definecolor{limegreen}{rgb}{0.2,0.7,0.2}
\definecolor{navyblue}{RGB}{0,0,128}
\definecolor{aquamarine}{RGB}{102,205,170}

\definecolor{strictRED}{RGB}{184,0,0}
\definecolor{specificationTURQUOISE}{RGB}{0,128,153}
\definecolor{assumptionGREEN}{RGB}{0,128,0}
\definecolor{interruptBLUE}{RGB}{0,0,128}
\definecolor{committedORCHID}{RGB}{54,22,89}
\definecolor{urgentORCHID}{RGB}{74,28,109}
\definecolor{requestedORCHID}{RGB}{104,34,139}
\definecolor{eventuallyORCHID}{RGB}{154,50,205}

\newboolean{showtext}
\setboolean{showtext}{false}  

\newcommand{\change}[1]{%
  \ifthenelse{\boolean{showtext}}{\textcolor{red}{#1}}{}
  }

\lstdefinelanguage{SMLX}
{
	basicstyle=\ssmall\ttfamily, %
	frame=single, 
	framextopmargin=0pt,
	framexbottommargin=0pt,
	framexleftmargin=0pt,
	xleftmargin=16pt,
	xrightmargin=3pt,
	morekeywords=[1]{system, domain, scenario, bind, to, 
		message, non, spontaneous, events, specification, 
		alternative, if, collaboration, role, with, dynamic, 
		bindings, or, and, null, define, as, 
		constraints, import, static, parameter, ranges, var, EInt, 
		controllable},
	morekeywords=[2]{strict},
	morekeywords=[3]{forbidden, violation},
	morekeywords=[4]{interrupt},
	morekeywords=[5]{guarantee},
	morekeywords=[6]{assumption}, 
	morekeywords=[7]{committed}, 
	morekeywords=[8]{urgent},
	morekeywords=[9]{requested},
	morekeywords=[10]{eventually},
	keywordstyle=[1]\color{darkviolet}\textbf,
	keywordstyle=[2]\color{strictRED}\textit,
	keywordstyle=[3]\color{strictRED}\textit,
	keywordstyle=[4]\color{interruptBLUE}\textit,
	keywordstyle=[5]\color{specificationTURQUOISE}\textbf,
	keywordstyle=[6]\color{assumptionGREEN}\textbf,
	keywordstyle=[7]\color{committedORCHID}\textit,
	keywordstyle=[8]\color{urgentORCHID}\textit,
	keywordstyle=[9]\color{requestedORCHID}\textit,
	keywordstyle=[10]\color{eventuallyORCHID}\textit,
	sensitive=false,
	morecomment=[l][\color{darkgreen}\textit]{//},
	morecomment=[s][\color{darkgreen}\textit]{/*}{*/}, 
	morestring=[b][\color{blue}]",
	tabsize=1,
	moredelim = [s][\color{specificationTURQUOISE}\textbf]{guarantee}{scenario},
	moredelim = [s][\color{assumptionGREEN}\textbf]{assumption}{scenario},
	backgroundcolor=\color{lightgray}
}

\lstdefinestyle{SMLXStyle} {language=SMLX}

\lstdefinelanguage{SMLConfig}
{
	basicstyle=\ssmall\ttfamily,
	frame=single, 
	framextopmargin=0pt,
	framexbottommargin=0pt,
	framexleftmargin=0pt,
	xleftmargin=16pt,
	xrightmargin=3pt,
	morekeywords=[1]{symbolic, import, configure, specification, use, 
		instancemodel, symbolic, parameters, attributes, symbolic, state, matching, 
		off, under, approximation, on, rolebindings, collaboration, object, plays, 
		role, role1},
	keywordstyle=[1]\color{darkviolet}\textbf,
	sensitive=false,
	morecomment=[l][\color{darkgreen}\textit]{//},
	morecomment=[s][\color{darkgreen}\textit]{/*}{*/}, 
	morestring=[b][\color{navyblue}\textit]",
	stringstyle=\color{navyblue},
	tabsize=1,
	backgroundcolor=\color{lightgray}
}

\lstdefinestyle{SMLConfigStyle} {language=SMLConfig}
\lstdefinelanguage{Java}
{
	basicstyle=\ssmall\ttfamily,
	frame=single, 
	framextopmargin=0pt,
	framexbottommargin=0pt,
	framexleftmargin=0pt,
	xleftmargin=16pt,
	xrightmargin=3pt,
	morekeywords=[1]{public, private, class, extends, protected, void,
		new, throws, null, if, else},
	morekeywords=[2]{STRICT},
	morekeywords=[3]{@Override},
	morekeywords=[4]{car, oc, cp}, 
	keywordstyle=[1]\color{darkviolet}\textbf,
	keywordstyle=[2]\color{javablue}\textbf,
	keywordstyle=[3]\color{darkgray},
	keywordstyle=[4]\color{navyblue},
	sensitive=false,
	morecomment=[l][\color{javagreen}\textit]{//},
	morecomment=[s][\color{javagreen}\textit]{/*}{*/}, 
	morestring=[b][\color{javablue}\textit]",
	stringstyle=\color{navyblue},
	tabsize=1,
	backgroundcolor=\color{lightgray}
}

\definecolor{javared}{rgb}{0.6,0,0} 
\definecolor{javablue}{rgb}{0,0,0.9} 
\definecolor{javagreen}{rgb}{0.25,0.5,0.35} 
\definecolor{javapurple}{rgb}{0.5,0,0.35} 
\definecolor{javadocblue}{rgb}{0.25,0.35,0.75} 

\lstdefinestyle{JavaStyle} {language=Java}

\lstdefinelanguage{Kotlin}{
	basicstyle=\ssmall\ttfamily,
	frame=single, 
	framextopmargin=0pt,
	framexbottommargin=0pt,
	framexleftmargin=0pt,
	xleftmargin=16pt,
	xrightmargin=3pt,
	comment=[l]{//},
	commentstyle={\color{darkgray}\ttfamily},
	emph={delegate, filter, first, firstOrNull, forEach, lazy, map, mapNotNull, println, return@, event, sends, request, requestParamValuesMightVary},
	emphstyle={\color{darkviolet}},
	identifierstyle=\color{black},
	numberstyle=\color{darkgreen},
	keywords=[1]{ abstract, actual, as, as?, break, by, companion, continue, data, do, dynamic, else, enum, expect, false, final, for, get, if, import, in, interface, internal, is, null, object, override, package, private, public, return, set, super, suspend, this, throw, true, try, typealias, val, var, vararg, when, where, while},
	keywordstyle=[1]{\color{javablue}\bfseries},
	keywords=[2]{@Deprecated, @JvmField, @JvmName, @JvmOverloads, @JvmStatic, @JvmSynthetic, @Test, Array, Byte, Double, Float, Int, Integer, Iterable, Long, Short, String, scenario, class, fun},
	keywordstyle=[2]{\color{javablue}},	
	keywords=[3]{interruptingEvents, forbiddenEvents, it}, 
	keywordstyle=[3]{\color{darkviolet}\bfseries},
	keywords=[4]{scenario, cycleScenario, runTest, Given, When, Then, And, But}, %
	keywordstyle=[4]{\textit},
	keywords=[5]{coolantTemp, deratingFactor}, 
	keywordstyle=[5]{\color{darkviolet}\bfseries\underbar},
	keywords=[6]{currentTemp}, 
	keywordstyle=[6]{\underbar},
	morecomment=[s]{/*}{*/},
	morecomment=[s][\color{black}]{`}{`},
	morestring=[b]",
	morestring=[s]{"""*}{*"""},
	sensitive=true,
	stringstyle={\color{javagreen}\ttfamily},
}

\lstdefinestyle{KotlinStyle} {language=Kotlin}

\newcommand{\lstinlineKotlin}[1]{\lstinline[language=Kotlin,basicstyle=\ttfamily]{#1}}

\ulposdef{\ulnumaux}{%
   $\underset{\saveulnum}{\rule[-.7ex]{\ulwidth}{.4pt}}$}

\articletype{Regular Article}%

\received{Date Month Year}
\revised{Date Month Year}
\accepted{Date Month Year}
\journal{Journal}
\volume{00}
\copyyear{2023}
\startpage{1}

\begin{document}

\title{Model-based Analysis and Specification of Functional Requirements and Tests for Complex Automotive Systems}

\author[1]{Carsten Wiecher}

\author[2]{Constantin Mandel}

\author[3]{Matthias Günther}

\author[4,5]{Jannik Fischbach}

\author[6]{Joel Greenyer}

\author[7]{Matthias Greinert}

\author[8]{Carsten Wolff}

\author[3]{Roman Dumitrescu}

\author[9]{Daniel Mendez}

\author[2]{Albert Albers}

\authormark{Wiecher \textsc{et al.}}
\titlemark{Model-based Analysis and Specification of Functional
Requirements and Tests}

\address[1]{\orgname{Kostal Automobil Elektrik GmbH \& Co. KG}, \orgaddress{\country{44227 Dortmund, Germany}}}

\address[2]{\orgname{IPEK – Institute of Product Engineering at Karlsruhe Institute of Technology (KIT)}, \orgaddress{\country{76131 Karlsruhe, Germany}}}

\address[3]{\orgname{Fraunhofer IEM}, \orgaddress{\country{33102 Paderborn, Germany}}}

\address[4]{\orgname{fortiss GmbH}, \orgaddress{\country{80805 Munich, Germany}}}

\address[5]{\orgname{Netlight Consulting GmbH}, \orgaddress{\country{80538 Munich, Germany}}}

\address[6]{\orgname{FHDW Hannover}, \orgaddress{\country{30173 Hannover, Germany}}}

\address[7]{\orgname{Two Pillars GmbH}, \orgaddress{\country{33102 Paderborn, Germany}}}

\address[8]{\orgname{FH Dortmund}, \orgaddress{\country{44139 Dortmund, Germany}}}

\address[9]{\orgname{Blekinge Institute of Technology,}, \orgaddress{\country{371 41 Karlskrona, Sweden}}}

\corres{Corresponding author Carsten Wiecher,  \email{c.wiecher@kostal.com}}


\fundingInfo{the German Federal Ministry of Education and Research (BMBF) within the “The Future of Value Creation – Research on Production, Services and Work” program (funding number 02J19B106 - 02J19B090) as part of the research project MoSyS - Human - Oriented Design of Complex Systems of Systems, which is managed by the Project Management Agency Karlsruhe (PTKA). The authors are responsible for the content of this publication.}

\abstract[Abstract]{The specification of requirements and tests are crucial activities in automotive development projects. However, due to the increasing complexity of automotive systems, practitioners fail to specify requirements and tests for distributed and evolving systems with complex interactions when following traditional development processes. 
To address this research gap, we propose a technique that starts with the early identification of validation concerns from a stakeholder perspective, which we use to systematically design tests that drive a scenario-based modeling and analysis of system requirements. 
To ensure complete and consistent requirements and test specifications in a form that is required in automotive development projects, we develop a Model-Based Systems Engineering (MBSE) methodology. This methodology supports system architects and test designers in the collaborative application of our technique and in maintaining a central system model, in order to automatically derive the required specifications. 
We evaluate our methodology by applying it at KOSTAL (Tier1  supplier) and within student projects as part of the masters program Embedded Systems Engineering. Our study corroborates that 
our methodology is applicable and improves existing requirements and test specification processes by supporting the integrated and stakeholder-focused modeling of product and validation systems, where the early definition of stakeholder and validation concerns fosters a problem-oriented, iterative and test-driven requirements modeling. 
}


\maketitle



\section{Introduction}

\textbf{Motivation}
Connected vehicle systems communicate with changing and evolving external systems, such as different types of charging infrastructures or user devices \cite{Kirpes2019}. This leads to continuously changing requirements. For example, a vehicle functionality may need to be synchronized with feature updates of a smartphone app \cite{Vogelsang2020}. For companies in complex automotive development partnerships this leads to the challenge that frequently changing \textit{stakeholder requirements}, which express business needs or business goals, must be compiled into high-quality \textit{system requirements} and \textit{system tests} in an iterative system development \cite{Kasauli2021}.
The decomposition of stakeholder requirements to system requirements is part of the requirements analysis phase as defined in automotive standards \cite{AutomotiveSIG2015, Mueller2016, ISO26262_2018}, where the resulting artifact is the \emph{system requirements specification}. 
This specification is a key boundary document \cite{Wohlrab2020} for subsequent system design and validation activities. The validation activities in turn rely on \emph{test specifications} as a result of the test design phase \cite{AutomotiveSIG2015, Juhnke2020}. 
Consequently, system requirements and test specifications determine the subsequent system design and validation, and hence the overall system quality. Moreover, they have a high impact on the collaborative development of automotive systems across departments and companies \cite{Kasauli2021, Wohlrab2020}. 

\textbf{Problem and Previous Work}
Due to frequent changing stakeholder requirements, practitioners increasingly fail to specify high-quality system requirements and tests for complex automotive systems \cite{Kasauli2021, Liebel2019}. We already addressed this research gap in previous work and proposed an \emph{integrated and iterative requirements analysis and test specification} approach to support both the analysis and specification of requirements and tests in an early development phase \cite{Fischbach2022, Wiecher2019, Wiecher2020, Wiecher2021, Wiecher2021c} 


Motivated by the test-driven development paradigm \cite{Beck2003}, this approach is based on the idea that requirements modeling can be driven by test cases to create immediate feedback in short iterations and increase the applicability of formal requirements modeling in practice. Instead of deriving test cases from system requirements manually, we use existing stakeholder requirements to generate acceptance tests that drive the modeling of system requirements. We introduced this technique as \emph{test-driven scenario specification} (TDSS) \cite{Wiecher2019} 

We found that TDSS can increase requirements quality by automatically identifying contradictions in automotive software requirements in an early development phase \cite{Wiecher2019}. We also found that the automated test case design improves the existing manual test case design process \cite{Wiecher2021c, Fischbach2022}. In addition, we showed that the test case design and TDSS can be combined in a way that the resulting requirements model is suitable to generate UML sequence diagrams to automatically specify functional system requirements \cite{Wiecher2021c}. Consequently, our technique increases the quality of test and requirements specifications and is suitable to document complex system interactions already in the requirements analysis phase \cite{Wiecher2019, Wiecher2021c, Fischbach2022}. 

\textbf{Scope of this Work}
In our preceding work we considered an isolated development context with a limited set of automotive system and component requirements. In practice, however, development iterations including the definition and analysis of system requirements are triggered continuously by changing stakeholder requirements \cite{Kasauli2021}. Consequently, since requirements and test specifications act as communication interfaces for the product development \cite{Wohlrab2020}, it is necessary to also consider different stakeholders, their concerns, and the resulting impact on the system requirements. 

According to Albers et al. \cite{Albers2010, AlbersBehrendt2017}, validation is the connecting activity in product development to align stakeholder concerns and the system in development as well as to gain new knowledge about the system, which must be recorded in the form of system requirements. Consequently, validation should be performed continuously throughout the development process and especially in the early development phase \cite{Mandel2021}. 

Following this understanding of early and continuous validation, we use a Model-Based Systems Engineering (MBSE) methodology that starts with the early definition of \emph{validation concerns} based on existing stakeholder requirements and use-cases.
In combination with our TDSS technique, this allows us to verify if each stakeholder requirement is modeled sufficiently by one or more system requirements, and to validate if the resulting system requirements specification fulfills the identified validation and stakeholder concerns respectively. In this way, the specification of system requirements is continuously driven by validation concerns, and the current knowledge is documented in complete and consistent requirements and test specifications, that can be derived from a central system model automatically. 

 
\textbf{Contributions}
Following the design science research (DSR) paradigm  \cite{Hevner2004,VomBrocke2020} and based on our previous work \cite{Wiecher2021c, Fischbach2022, Mandel2021}, our results are composed as a new DSR artifact which we applied at a Tier-1 supplier company. We make the following contributions: 
\begin{itemize}
    \item As a first step, we identified relevant system elements necessary to integrate the TDSS technique into an MBSE process. Therefore, we focused on the integration of stakeholder concerns and related stakeholder requirements to express business needs and goals. 
    In addition, we found that existing MBSE approaches (e.g., Holt \& Perry \cite{HoltPerry2019}) must be extended to consider validation concerns and their dependencies to the system in development sufficiently (cf. Mandel et al. \cite{Mandel2021}). 
    To address this, we propose an ontology (see Fig. \ref{fig:ontology}) that integrates the concepts presented in \cite{Mandel2021} for a comprehensive and continuous validation system modeling.  
    In this way, the ontology supports both the modeling of stakeholder concerns and the modeling of distributed validation environments as used in the automotive industry. Consequently, we ensure that the resulting test specification is suitable to validate the identified stakeholder requirements.  
    
    \item Second, based on our ontology, we consolidated specific viewpoints and views to handle the system complexity and support system architects and test designers in the focused and collaborative system modeling, by editing specific subsets of the system model depending on the available information and situation within the development project. 
    The concrete modeling tasks are defined with a set of activities that integrate with our test-driven modeling technique (see Fig. \ref{fig:mbsescil}). 
    The results are implemented as an MBSE framework with an MBSE tool.\change{C2}
    
    
    \item Finally, we evaluate our modeling and analysis technique with the help of complex functionality of an on-board charger (OBC) control unit \cite{Schnitzler2020} for battery electric vehicles, to investigate if the resulting specifications fulfill the demands of automotive development projects. 
    We demonstrate that the MBSE methodology is suitable to process real-world stakeholder requirements. The proposed method (Fig. \ref{fig:mbsescil}) supports an early structuring of stakeholder requirements, which enables the definition of validation concerns already in the requirements analysis phase. 
    We found that this explicit and early modeling of validation concerns in combination with TDSS is suitable to provide an orientation for the modeling, analysis, and documentation task. 
    Compared to the established development process at the Tier-1 supplier company, we identified that our technique is beneficial to plan distributed validation tasks and align these tasks to specific validation goals.   
    Our validation-focused MBSE technique improves the state of practice for the specification of system requirements and test. \change{C3}
\end{itemize}

\textbf{Outline}
This paper is structured as follows: In Sect. \ref{sect:context} we introduce the relevant context. In particular, we introduce the demands for requirements and test specifications in automotive development projects. Sect. \ref{sect:foundations} introduces our TDSS technique and related work on test-driven modeling and MBSE techniques. The developed MBSE methodology including the ontology, viewpoint and views, and the specification method are shown in Sect. \ref{sect:systemsmodeling}. The evaluation is presented in Sect. \ref{sect:casestudy}. We conclude in Sect.~\ref{sect:closing}.

\section{Context}
\label{sect:context}
\subsection{Automotive ECU Development}
On a macro-level, automotive ECU development is structured according to the V-model \cite{VDI2004} and the quality of the resulting artifacts is measured in stage-gate management processes \cite{Cooper1994} to ensure that the developed systems comply with high quality and safety demands. In practice, however, it is increasingly difficult to follow this plan-driven development approach \cite{KasauliEASE2020}, since safety critical automotive systems are increasingly interacting with external systems (e.g, charging stations, smart home infrastructure) to provide an integrated functionality \cite{Kirpes2019, Vogelsang2020}. This is leading to frequent changing stakeholder needs and related requirements updates, since the functionality of the automotive system must be aligned with other systems developed independently in other organizations  \cite{Bohm2021}.   
As a result, automotive companies try to adopt agile methods into plan-driven development environments to decrease the response times on new stakeholder needs  \cite{KasauliEASE2020}. However, recent studies have shown that particularly the specification of requirements and tests within a more iterative development process is challenging, since the analysis and specification are primarily carried out manually. Combined with frequent change requests from various stakeholders, it becomes increasingly difficult to provide complete and consistent specifications in the development of complex systems \cite{Kuhrmann2021, KasauliEASE2020}. 

\subsection{Requirements Specification}
\label{sect:requirementspecification}
The state of practice for the specification of ECU requirements is based on top-down requirements decomposition approaches~\cite{Penzenstadler2010,Bohm2014}, 
where stakeholder requirements are decomposed to system requirements that in turn are linked and decomposed to subsystem and component requirements. 
This corresponds to the system life cycle process defined in ISO 15288 for systems and software engineering \cite{ISO/IEC/IEEE2015}. 

Accordingly, \emph{stakeholder concerns} can be seen as a starting point for the development task. They comprise needs, wants, desires, expectations and preconceived constraints of identified stakeholders \cite{ISO/IEC/IEEE2015}. These concerns are determined from the communication with external and internal stakeholders \cite{INCOSE2015}. 

Based on the stakeholder concerns and related stakeholder requirements specifications, the focus of this article is on the specification and analysis of functional system requirements that must be collected in a \emph{system requirements specification} \cite{ISO/IEC/IEEE2015}\cite{INCOSE2015}.  
The specification and analysis of system requirements is part of automotive development processes. These processes must include a requirements analysis phase with the purpose to "(...) transform the defined stakeholder requirements into a set of system requirements that will guide the design of the system" \cite{AutomotiveSIG2015}. As an outcome of this process, it is expected that a system requirements specification is established, where the containing requirements are analyzed for correctness and verifiability, and that consistency and bidirectional traceablity between system and stakeholder requirements are ensured~\cite{AutomotiveSIG2015}. 


Although many different attempts have been made to support this early requirements analysis phase through formal and model-based techniques (e.g., Holtmann \cite{Holtmann2019} \& Greenyer et al. \cite{Greenyer2015}), the analysis of functional requirements is still a predominantly manual process~\cite{Liebel2019}. 

\subsection{Test Specification}
\label{sect:testspecification}
In automotive development projects, requirements-based testing is the predominant activity to ensure that a \emph{test object} (e.g., ECU, or intermediate stages of partially integrated systems) is compliant with a \emph{test basis} (e.g., system requirements specification)~\cite{Juhnke2020, Spillner2011}.   

The key boundary document for the testing activities is the \emph{test specification}, which acts as a communication interface between different roles across companies and departments. This test specification is a complete documentation of the test design, including test cases and the test procedures for a specific test object \cite{ISO/IEC/IEEE2013}.

The specification itself is created by test designers. Subsequently, testers can be seen as consumers that use the test specification as a basis for the implementation of test automation scripts, or to perform manual tests \cite{Juhnke2020}. Thereby, the implementation highly depends on specific \emph{test environments} which must be constantly aligned with the development progress of the test object \cite{Mandel2021}. In this article, the term test environment is used to describe hardware, instrumentations, simulators, software tools, and other support elements needed to conduct a test \cite{Spillner2011, AlbersBehrendt2017}. 


Based on a given test environment, it is necessary that the test cases are suitable to test the interaction of system items and that the results are recorded, where a consistency and bidirectional traceability between the system architectural design and the test cases, and between the test cases and test results is established \cite{AutomotiveSIG2015}. 

\subsection{Motivating Example}
\label{sect:example}
As outlined in \cite{Kirpes2019}, e-mobility systems will be composed by a large number of battery electric vehicles, smart charging stations, and information systems that interlink the electricity and mobility sector. To describe these emerging systems, a system of systems (SoS) perspective can be suitable \cite{Keating2008}, where systems are developed independently in different organizations, but provide an integrated functionality. 

For the development of single systems and their subsystems (e.g, battery electric vehicles, ECUs), this SoS context inclu\-ding evolving interacting systems is especially challenging for the previously introduced requirements analysis and test specification phases of automotive development projects \cite{Ncube2018, Dumitrescu2021, Honour2013}. 
The system developer (Tier 1 supplier in the context of this work) must be supported in the modeling tasks to iteratively respond to changing stakeholder requirements and deliver complete and consistent specifications.

\begin{figure}[h]
    \centering
    \includegraphics[width=1.0\linewidth]{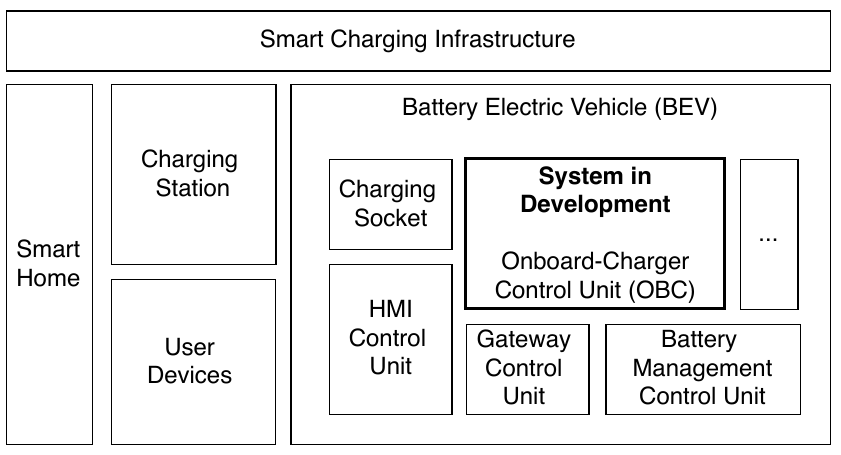}
    \caption{OBC as a system in development in an e-mobility system of systems context.}
    \label{fig:example}
\end{figure}

E.g, Fig. \ref{fig:example} outlines a battery electric vehicle including several ECUs required to realize a charging functionality in an SoS context.  
One central ECU is the OBC that converts AC voltage from the grid into DC voltage of the vehicle's battery \cite{Schnitzler2020}. To satisfy the needs of a global market, it is necessary to support efficient charging at different power grids in different regions (e.g., three-phase 11kW grids in Europe, or single phase charging at 7.2kW in China). In addition, different charging standards (e.g., CHAdeMO \cite{CHAdeMOassociation2012}, Chinese GB/T DC standard~\cite{GB/T20234}) must be supported. 

This requires that a high technical complexity must be considered in the requirements analysis phase, and that test specifications exists that are suitable to automate tests on validation environments that consider these different types of hardware and software environments and standards. 

Based on this technical background, we use the concrete function \emph{timer-charging} to evaluate our approach in Sect. \ref{sect:casestudy}. This function extends the basic charging functionality of the vehicle to allow users to configure properties like the expected state of charge (SOC) that shall be reached at a specific time. Therefore, the OBC of the battery electric vehicle communicates with internal (HMI control unit) or external (smartphone app) user interfaces and processes the user input depending on the current charging environment.

\section{Foundations \& Related Work}
\label{sect:foundations}

\subsection{Test-Driven Modeling}
Originally introduced in software development, test-driven development aims to shift quality assurance from reactive to proactive work to support the development of reliable software in a short iterations \cite{Beck2003}. 
Following the red/green/refactor mantra, the implementation of a new functionality starts with writing a test which is expected to fail initially or which even doesn´t compile (red). The second step covers everything what needs to be done to get this test working (green), which can be adapted in the third step, e.g.,  
by deleting overlaps caused by the mere execution of the test (refactor) \cite{Beck2003}.    

Based on positive findings in the application of test-driven development, e.g., with respect to error rate, productivity and test frequency \cite{Maximilien2003}, several approaches emerged to use this principle also for modeling tasks \cite{Zugal2012,Zhang2004,glinz2007}. 

Although test-driven development in combination with requirements modeling seems to be a promising approach to support the requirements analysis in automotive development projects, the related work \cite{Zhang2004, glinz2007} doesn't consider a systematic specification of tests to be used to drive the modeling. In addition, since these contributions rely on scenario-based modeling, addressing the sampling and coverage concern \cite{Sutcliffe2003} is necessary but not covered sufficiently.

Consequently, we integrate the systematic test case specification and scenario-based modeling technique introduced in our previous work \cite{Wiecher2021c, Fischbach2022} with the newly developed MBSE technique in this project. This supports the system architect and test designer to identify when a representative set of requirements is modeled for a current problem / stakeholder concern (sampling), and if an adequate set of test scenarios exists for requirements validation (coverage). \change{C4} \change{C5}

\subsection{Test-Driven Scenario Specification (TDSS)} 
\label{sect:smlk}
To handle complexity and to foster a common system understanding, scenarios are seen as beneficial for requirements engineering and system validation tasks \cite{Sutcliffe2003, Kaner2003}. Therefore, in this work, we use scenario-based techniques for systems and requirements modeling. For systems modeling, we use scenarios to decompose use-cases, structure stakeholder requirements, and identify stakeholder and validation concerns. For requirements modeling we apply the scenario modeling language for Kotlin (cf. Greenyer \cite{Greenyer2021}) to automatically analyze and document functional system requirements.  


To support practitioners in the application of this formal requirements modeling task, we introduced the TDSS technique \cite{Wiecher2019} to iteratively model single requirements driven by tests, where the ability to immediately test the modeled requirements brings high confidence and a feeling of control to the requirements specification and analysis phase \cite{Wiecher2019}.
\begin{figure}[h]
    \centering
    \includegraphics[width=1.0\linewidth]{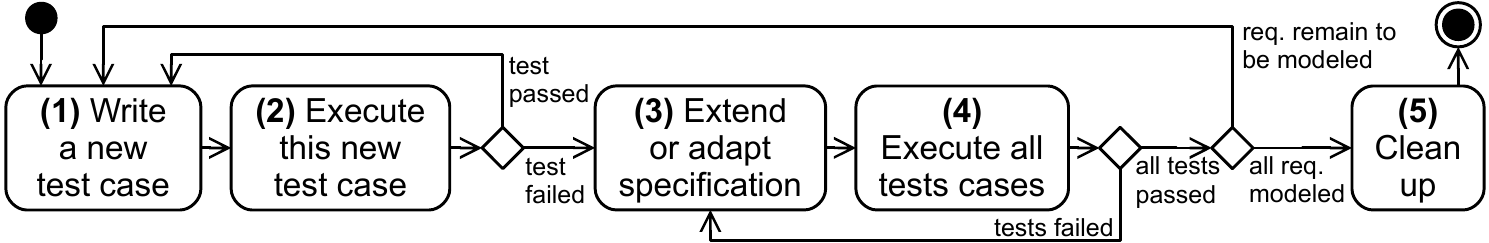}
    \caption{Test-driven specification of system requirements using the scenario-based modeling language for Kotlin. \cite{Wiecher2019}}
    \label{fig:tdss}
\end{figure}

The TDSS process is illustrated in Figure \ref{fig:tdss}. It starts with writing a new test case, which is then executed in the second step. If this test case fails, the formal specification of the system requirements is to be adjusted in the third step. Subsequently, the new test case, along with all test cases from previous iterations, is executed. This clarifies whether the new system requirement has been correctly formalized and whether there are any functional dependencies with existing requirements. If all test cases are successfully executed in this step, the process concludes with cleaning up the created specifications.

\subsection{Model-Based Systems Engineering (MBSE)}
\label{sect:Model-Based Systems Engineering (MBSE)}
MBSE is regarded as a promising approach to managing system complexity and overcoming challenges in product engineering.\cite{INCOSE2021}. As described by Walden et al., MBSE  is the formalized application of modeling to support engineering activities such as verification and validation over the whole product lifecycle \cite{INCOSE2015}. The goal of MBSE is to replace the multitude of independent documents used in engineering (e.g. system description documents, requirements and test specifications etc.) by a central system model \cite{INCOSE2015}. In order to develop system models, different methods and methodologies can be used \cite{estefan2007,weilkiens2016,Douglass2017}. They often comprise a process model and activities to achieve a certain goal. The methods can be divided into generic methods, but also language- or tool-specific methods (cf. Weilkiens et al. \cite{weilkiens2016}).
Although extensive work has been done in the field of architectural modeling in MBSE, the respondents of various studies still see validation and verification as a major challenge that has not been covered comprehensively \cite{Dumitrescu2021, Cloutier2019}.

As mentioned by Husung et al. \cite{husung2021}, different use cases, such as the definition of verification and validation criteria should be supported using systems modeling. Currently only a few approaches, such as the SPES XT modeling framework \cite{Pohl2016} or the object-process framework by Langford \cite{langford2017}, include uses cases for verification and validation in MBSE modeling.
While the SPES XT modeling framework provides a generic input on how to support validation activities in an architecture model, the framework does not support specific artifacts and viewpoints for the continuous support of verification and validation. The object-process framework \cite{langford2017} on the other hand focuses on process-related aspects without providing an ontology to formalize information. Another approach that addresses validation in systems modeling is proposed by Lindeman et al. \cite{Lindemann2016}. However, this Framework does not specify methods, viewpoints, guidelines etc. to build up and use a system model in the understanding of MBSE.

Although a few concepts exist, there is a need for action in the use of an integrated system model for early and continuous support for test planning and analysis in product development. Especially the detailed system information consolidated in a system model is useful in the area of validation planning. For example context information, as mentioned by Pohl et al. \cite{Pohl2016}, or the detailed interface specification that are specified in the structural models can be used to plan test environments. Therefore, in this paper we specifically consider verification and validation as part of an integrated MBSE approach, which is based on MBSE concepts as described by Holt \& Perry \cite{HoltPerry2019}. Accordingly, our MBSE approach consists of three aspects:  

\begin{itemize}
    \item \emph{Ontology}, for the definition of relevant terms and concepts used for modeling as well as their dependencies
    \item \emph{Viewpoints and Views}, focusing on specific subsets of the defined ontology and defining consistent sections thereof based on concerns of stakeholders of the MBSE approach
    \item \emph{(Architecture) Framework,} assembling and arranging relevant viewpoints and the ontology for a given application of the MBSE approach
\end{itemize}

In addition, a complementary method describing how to perform modeling based on the framework is required. This will ensure a consistent system model is created, which will enable full traceability capabilities across the architecture and validation artifacts.

\section{Model-based Analysis and Specification of Functional Requirements and Tests}
\label{sect:systemsmodeling}
Therefore, in this section, we present the three aspects: ontology, viewpoints and views, and a specification method for the application of the MBSE approach.
\subsection{Ontology for Test-Driven Analysis of Requirements}
\label{sect:ontology}
\begin{figure*}[ht]
    \centering
    \includegraphics[width=1.0\linewidth]{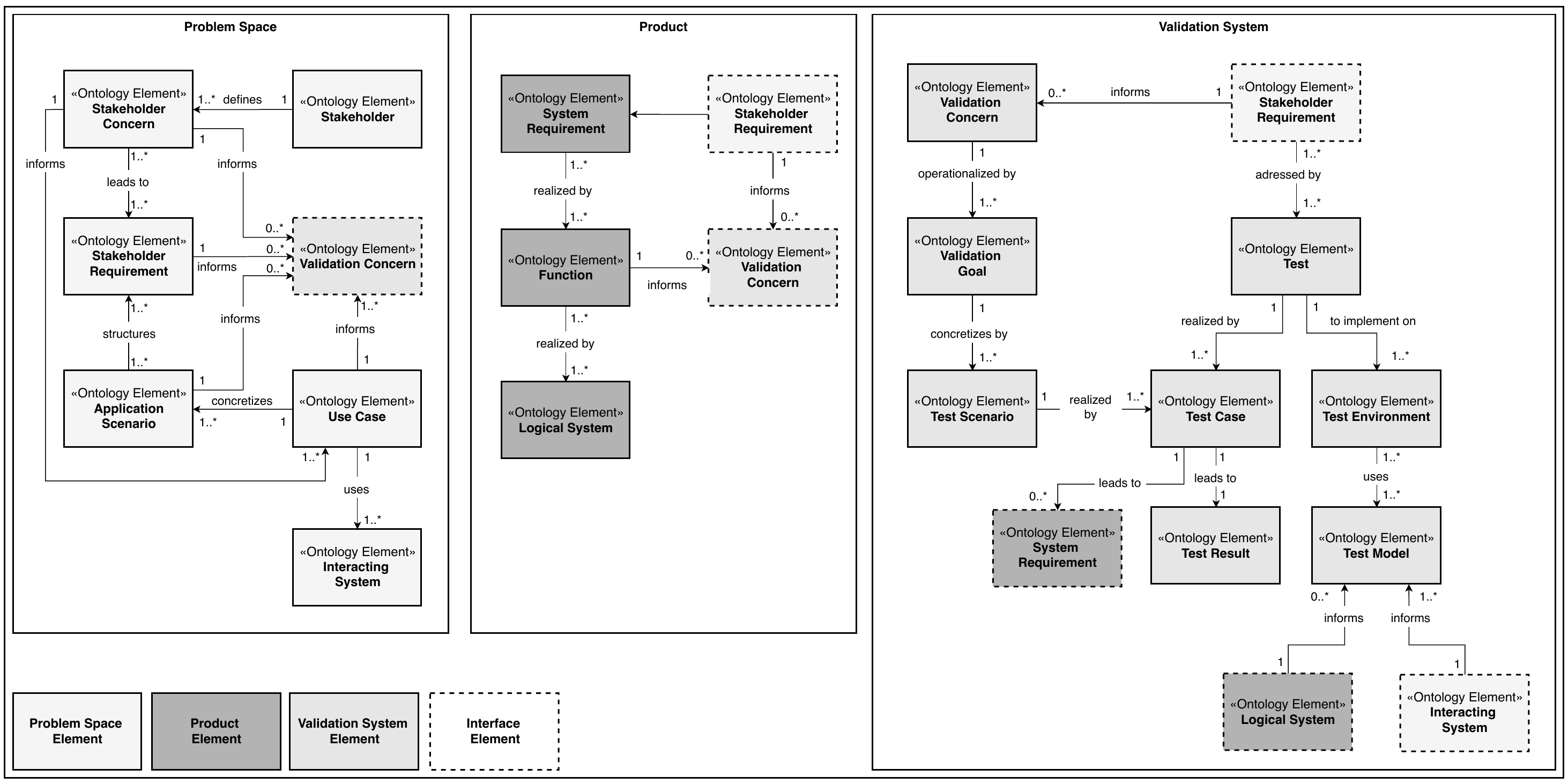}
    \caption{Excerpt from the ontology developed in the MoSyS research project.}
    \label{fig:ontology}
\end{figure*}
Holt \& Perry \cite{HoltPerry2019} describe a general MBSE ontology. However, the described ontology does not comprehensively enough cover elements for modeling of verification and validation aspects as needed for the goal of this work. Mandel et al. specifically investigate relevant elements for the description of verification and validation activities as well as their relations to define a (sub-) ontology for an MBSE approach \cite{Mandel2021}. This ontology is taken as a basis and is further specified in the BMBF-funded research project MoSyS. Relevant parts of the MoSyS ontology are used for the approach in this work. 

An excerpt of the used ontology is shown in Fig. \ref{fig:ontology}. We separate the single elements of the ontology in the sections \emph{Problem Space}, \emph{Product}, and \emph{Validation System}. 
The central part within the problem space are \emph{stakeholder requirements}, that are the starting point for sysetms engineering activities from a Tier1 supplier perspective and for our specification method (see Sect. \ref{sect:method}). These stakeholder requirements can be provided by external \emph{stakeholders} like OEMs, or internal stakeholders, e.g., functional safety, quality, and production managers, where the resulting stakeholder requirements are based on specific \emph{stakeholder concerns}. To support a context modeling, we consider \emph{interacting systems} from which, in interaction with the  \emph{stakeholders}, \emph{use cases} can be derived. The \emph{use cases} in turn can be concretized by \emph{application scenarios}. The information of the modeling within the problem space is linked to \emph{validation concerns} to model, where in the problem space knowledge gaps exists, that need to be resolved by appropriate validation activities. The dashed lines indicate that validation concerns are part of the validation system. 

The ontology elements for the product modeling are based on established MBSE approaches (e.g., \cite{estefan2007, HoltPerry2019, weilkiens2016, Pohl2016}), where \emph{system requirements} are derived from stakeholder requirements. The system requirements in turn are used to model the \emph{functional architecture} and eventually the \emph{logical architecture}. Within this article, we consider system requirements and functions as possible sources of validation concerns.   

The ontology elements of the validation system are essential for our test-driven modeling and analysis approach. The starting point within the validation space are the validation concerns, derived from the problem and solution space. The validation concerns are operationalized through \emph{validation goals}. To create a link from abstract validation goals to concrete \emph{test cases}, we use \emph{test scenarios} as a binding element. The element test in turn is a composition of \emph{test environment} and \emph{test case}. The test case leads to a \emph{test result} and the test environment is composed of \emph{test environment objects}, which can be derived from the logical architecture and the system context of the product and problem space respectively. 
This structure supports the alignment of modeling activities to conduct a test specification as required in automotive development projects, since the validation concern element is traceably linked to the elements from the problem space and product architecture. Accordingly, we can trace back the elements of the validation system based on their connection to the validation concern. This enables the implementation of specific viewpoints and views for test-driven modeling and analysis. \change{C13}

\subsection{Viewpoints and Views}
\label{sect:viewpoints}
Viewpoints can be understood as "filters" to show representations of a system mo\-del for sub-sets of the developed ontology (see Fig. \ref{fig:ontology}). In addition, each viewpoint is a blueprint to create one or more specific views in a (software) implementation of the model. The viewpoints/views are used to structure the system mo\-del. Each viewpoint can be used in a variety of mo\-deling activities of the specification method (see chapter \ref{sect:method}).
\begin{figure}[ht]
    \centering
    \includegraphics[width=1.0\linewidth]{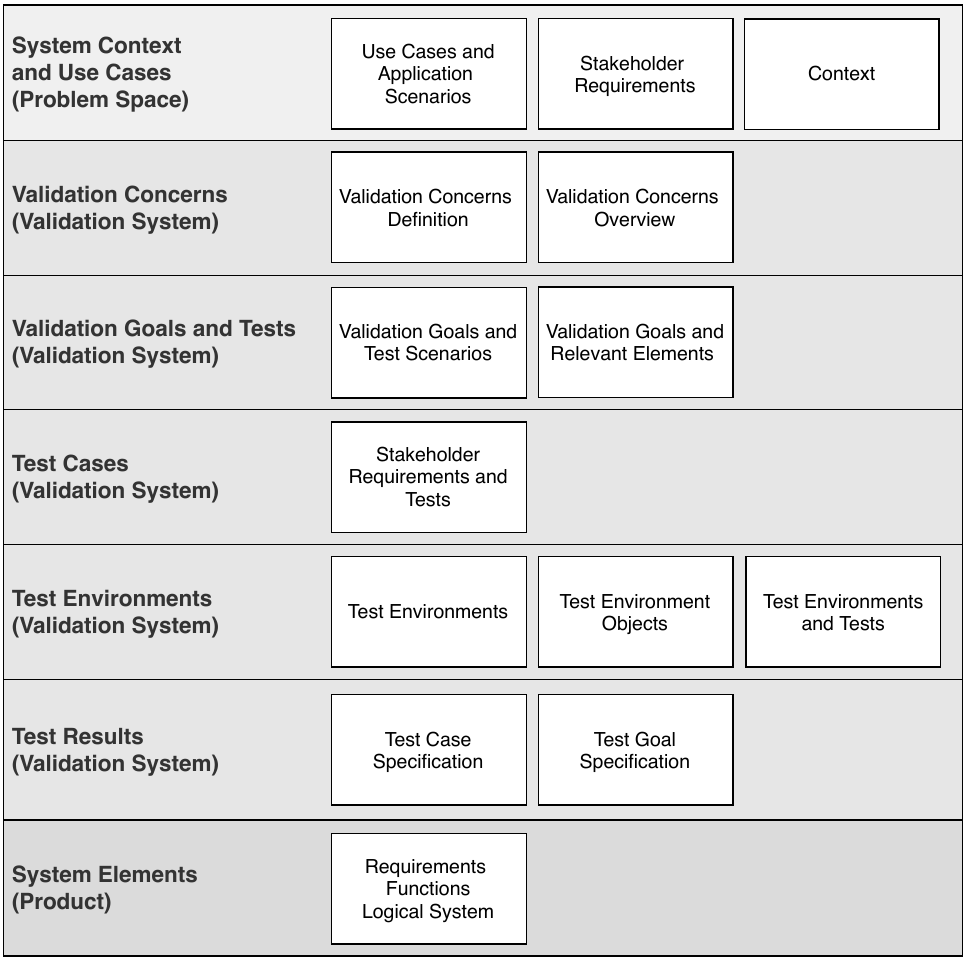}
    \caption{Viewpoints to describe the validation system.}
    \label{fig:viewpoints}
\end{figure}
The viewpoints we applied within this article focus on the validation system modeling. In addition, viewpoints for modeling the problem space and the product are included. The viewpoints are arranged in a framework of seven layers as shown in Fig. \ref{fig:viewpoints}, where the uppermost layer includes viewpoints for problem space modeling, the lowest layer covers viewpoints for product modeling and the remaining layers cover viewpoints for modeling of the validation system. Each layer contains one or more viewpoints:
\begin{itemize}
    \item \textbf{System Context and Use Cases (Problem Space)} supports the modeling of use cases and application scenarios from a stakeholder perspective, where each use case is based on a specific stakeholder concern. 
    In addition, the alignment of stakeholder requirements to the defined application scenarios and context modeling of the system in development is supported.
    Context systems interacting with the system in development via interfaces on the system boundary are modeled using the context viewpoint.
    
    \item \textbf{Validation Concerns (Validation System)} contains one viewpoint to define validation concerns based on the modeling activities within the problem space, and one viewpoint to provide an overview of all defined validation concerns including the addressed ontology elements.
    \item \textbf{Validation Goals and Tests (Validation System)} contains a viewpoint to concretize the list of validation concerns with specific validation goals and related test scenarios. In addition one viewpoint includes all relevant requirements and system elements that are related to a validation goal. 
    
    \item \textbf{Test Cases (Validation System)} contains a viewpoint to support the definition/generation of test cases based on given stakeholder requirements and defined validation goals. 
    
    \item \textbf{Test Environments (Validation System)} contains the viewpoints to define test environments and model the containing test environment objects. In addition this layer contains a viewpoint to support the assignment of tests to specific test environments. 

    \item \textbf{Test Results (Validation System)} contains viewpoints to derive the specifications for the implementation and execution of test cases and test scenarios for the requirements verification and validation. 
    
    \item \textbf{System Elements (Product)} is used to add system requirements, functions, and logical system elements to the system model, as a result of the test-driven modeling. 
\end{itemize}

\subsection{Specification Method} 
\label{sect:method}
\begin{figure*}[ht]
    \centering
    \includegraphics[width=0.8\linewidth]{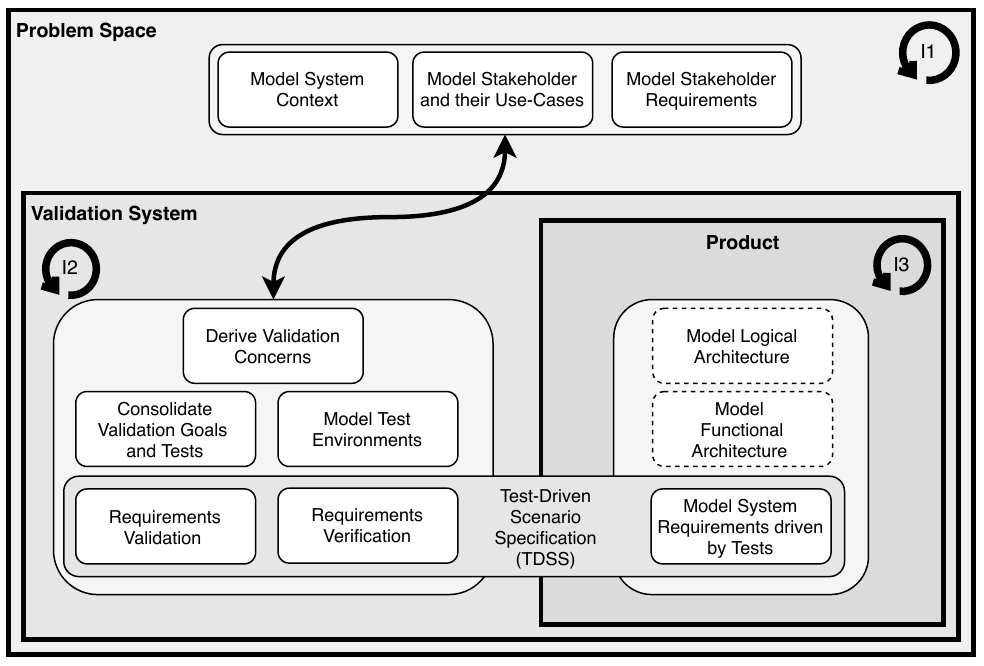}
    \caption{Integrated product and validation system modeling to obtain consistent and complete requirements and test specifications based on stakeholder and validation concerns.}
    \label{fig:mbsescil}
\end{figure*}
Fig. \ref{fig:mbsescil} presents an overview of the developed method, comprising eleven individually consistent modeling activities.
The goal of the method is to support system architects and test designers in the iterative and integrated development of complete and consistent requirements- and test specifications, that can be used as a basis for the detailed design of the product and the related validation systems. The modeling method integrates TDSS and the modeling method according to Mandel et al. \cite{Mandel2021} to support an early and continuous definition of test cases as well as the specification of requirements.
Based on the ontology (Sect. \ref{sect:ontology}) we propose three methodical blocks with different scopes: modeling of the \emph{problem space}, modeling of the \emph{validation system}, and modeling of the \emph{product}. 

The arrangement of these methodical blocks is motivated by two aspects. 
First, the system requirements and test specification tasks in automotive development projects are triggered by changing stakeholder concerns \cite{Kasauli2021}. Consequently, the subsequent activities for the validation system modeling in I2 are triggered and defined by changing \emph{system contexts}, \emph{use-cases}, and \emph{stakeholder requirements} captured in I1.
Second, the method builds on the idea of test-driven modeling \cite{Wiecher2019, Zugal2012} and integrates it with the concept of a continuous validation as a central development activity to create knowledge about the system in development \cite{Albers2010, AlbersBehrendt2017, Mandel2021}. Accordingly, we capture changing stakeholder information in I1, derive validation concerns in I2, and use this information to consolidate tests in order to drive the formal modeling of system requirements in I3. 

Specifically, in I2, we \emph{derive validation concerns} based on stakeholder use-cases and stakeholder requirements, but considering that system requirements and other information from previous development iterations may already exist, this information can also lead to validation concerns (see the dependencies in Fig. \ref{fig:ontology}). The defined validation concerns are the basis to \emph{consolidate validation goals and tests}. This activity supports the planning of validation tasks by addressing two challenges in complex development projects. First, what shall be validated in the upcoming development iteration? And second, where and with which resources should this be validated? In complex scenarios it is necessary to set a focus for the validation, which we do by assigning concrete validation goals to the list of all identified validation concerns. And since the validation activities are usually distributed across departments and companies, it is important to consolidate on which test environments which test cases should be executed. In this activity, this is supported by the ontology element \emph{test} that binds a \emph{test case} with a related \emph{test result} to a \emph{test environment} (see the validation system part of the ontology in Sect. \ref{sect:ontology}). If the existing test environments are not sufficient to fulfill the consolidated validation goal, the activity \emph{modeling of test environments} can be used to extend or add new test environments. 

These three activities in I2 are the foundation to execute the TDSS technique, in order to drive the modeling of system requirements and to automatically analyze if the specified system behavior corresponds to the stakeholder requirements. 
This is organized in a verification and validation path. In the verification path we execute the TDSS technique as introduced in Sect. \ref{sect:smlk}. As part of TDSS, we generate test cases for each stakeholder requirement that is relevant for the current development iteration. In this way, we ensure that each stakeholder requirement is verified by the right number of test cases (cf. \cite{Fischbach2022}) within the \emph{requirements verification} activity.
As a result of the activity \emph{model system requirements driven by tests}, we get a formal scenario-based requirements model. Subsequently, we use the validation path to investigate if the current validation goal is reached with the current implementation status of system requirements. For this \emph{requirements validation} we simulate the system requirements driven by test scenarios that are aligned to the validation goal (see Sect. \ref{sect:ontology}).\footnote{For details regarding the requirements modeling language please refer to \cite{Wiecher2019, Wiecher2021c, Greenyer2021}}. 

The execution of the individual modeling activities can be done sequentially from I1 to I3, but also in a different order depending on the currently available information and situation within the development project. Each modeling activity is linked to selected viewpoints from the overview in Fig.~\ref{fig:viewpoints} that support the analysis and syntheses for the modeling activity. On one hand, the use of iteratively reusable modeling activities can provide users with an intuitive and flexible means of modeling according to their specific problem situation. On the other hand, the ontology, viewpoints, and framework provide a structure for the created system models that can be reused across multiple projects and support users in navigating the models.

As result of the proposed approach, we iteratively add artifacts (a.o., test cases, test results, test scenarios, requirements) to our central system model, where the individual modeling activities are supported by the views and viewpoints introduced in Sect. \ref{sect:viewpoints}, and the single artifacts are related to each other according to the ontology introduced in \ref{sect:ontology}. As a key result of our approach, this allows us to derive complete and consistent requirements and test specification from the system model as shown in Fig. \ref{fig:testCaseSpecification} and Fig. \ref{fig:testGoalSpecification}. 

\section{Application and Evaluation}
\label{sect:casestudy}
\subsection{Overview}
\begin{figure}[ht]
    \centering
    \includegraphics[width=1.0\linewidth]{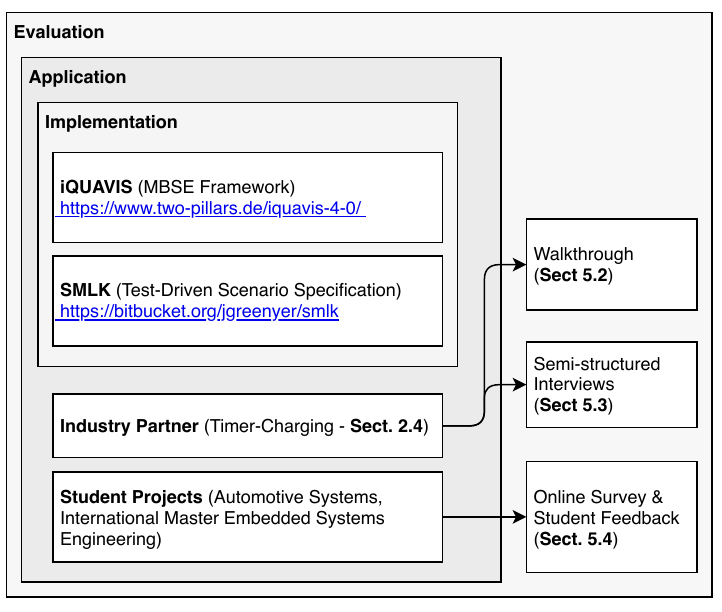}
    \caption{Implementation, application and evaluation of the proposed technique.}
    \label{fig:overviewevaluation}
\end{figure}

\begin{figure}[ht]
    \centering
    \includegraphics[width=1.0\linewidth]{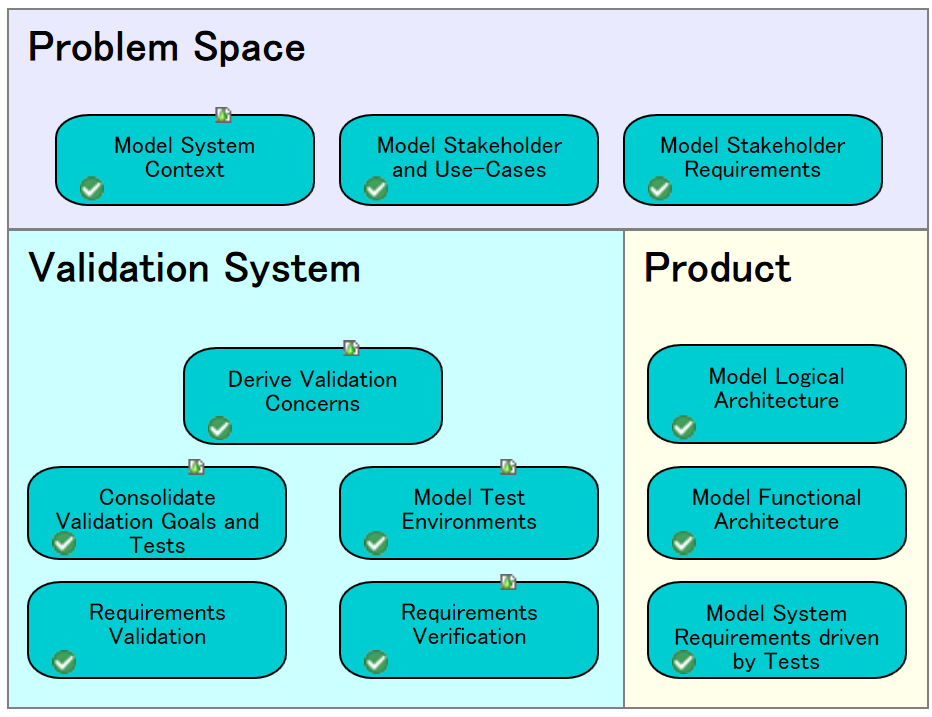}
    \caption{Implementation of the central entry point for modeling activities in iQUAVIS.}
    \label{fig:v&vframework}
\end{figure}

Fig. \ref{fig:overviewevaluation} shows the individual steps we conducted to evaluate our approach. Within the MoSyS research project, we implemented the ontology and the consolidated viewpoints and views with the MBSE tool iQUAVIS\footnote{\url{https://www.isidsea.co.th/iquavis}}. Subsequently, we built a modeling framework according to the concepts introduced in Sect.~\ref{sect:systemsmodeling}.
To support system architects and test designers in performing the proposed method (Sect. \ref{sect:method}), we implemented a central entry point from which all modeling activities can be accessed (see Fig. \ref{fig:v&vframework}).
Each activity contains links to the defined system views (see Fig. \ref{fig:viewpoints}) that are needed for the respective modeling task\footnote{Fig. \ref{fig:overviewevaluation} includes links to the used tools to reproduce our results. The tools for the requirements modeling and analysis are publicly available. iQUAVIS is not publicly available, but following the description in Sect. \ref{sect:systemsmodeling} and the proposed ontology in Fig. \ref{fig:ontology} other MBSE tools (e.g., Cameo Systems Modeler - cf. \cite{Mandel2021}) can be used to apply the proposed methodology. A list of additional tools can be found at: \url{https://mbse4u.com/sysml-tools} }. 

The evaluation is based on two elements:

First, we applied our approach at our industry partner. To make the application reproducible, we describe the individual steps in the form of a walk-through that follows the individual steps of our method. The results were discussed with experts at our industry partner and recorded with the help of an online survey.

Second, we applied our approach within student projects that took place in the summer term 2022 at \emph{dortmund university of applied sciences and arts} as part of the masters program \emph{embedded systems engineering}\footnote{\url{https://www.fh-dortmund.de/en/programs/embedded-systems-engineering-master.php}} (2nd semester). At the end of the student projects we conducted an online survey to record the applicability.  

The two elements of the evaluation highlight different aspects. The walk-through is the foundation to discuss our approach with experts at our industry partner. We used real-world requirements of a complex functionality to get reasonable feedback, e.g. on problem orientation and modeling depth. The walk-through was conducted by the authors of this work. In contrast, the student projects focus on smaller examples, but highlight the applicability of our approach by students with less MBSE experience. In this way, both the perspective of experts with extensive experience in system development and novices with less experience can be captured.

This evaluation framework was considered important because although experts in the industry provided valuable feedback based on the walk-through, their input relied on the documented application of the MBSE approach. The experts didn't independently devise the solutions. However, to ensure that the approach is applicable by third parties, the student projects and the feedback gathered there were crucial. This allows the relevance and applicability to be demonstrated. \change{C11}

\subsection{Walkthrough}
For the walk-through at our industry partner we started with a dataset of 67 stakeholder requirements. 
This set of stakeholder requirements was also the starting point for the development of the timer-charging function (introduced in Sect. \ref{sect:example}) following the established development process at the case company. 
The majority of the available requirements were functional requirements (57). We identified 4 architectural and quality requirements, e.g., the tolerance time of different clock values, or specific components that shall be considered in the architectural design were specified. In addition, 6 requirements specified concrete interfaces to components that were developed at external development partners.

\subsubsection{Problem Space Modeling}
\begin{figure}[ht]
    \centering
    \includegraphics[width=1.0\linewidth]{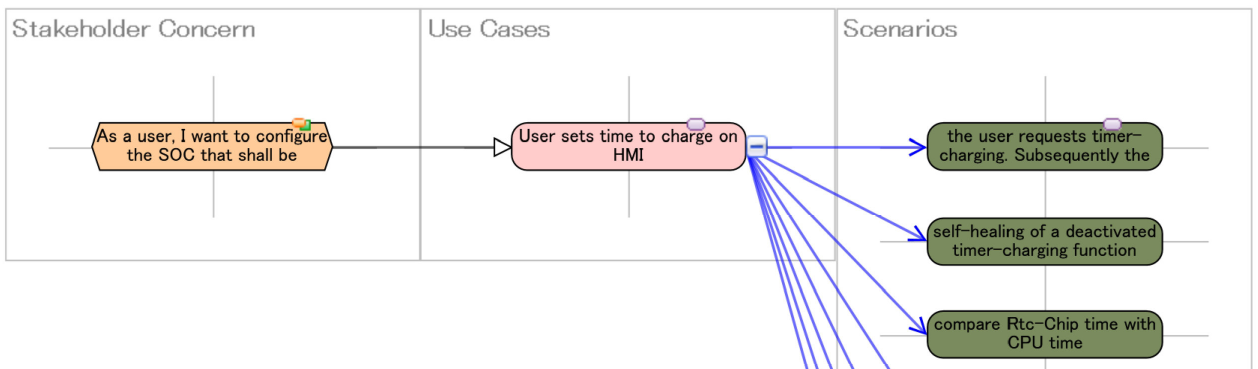}
    \caption{System view to define use cases and application scenarios based on stakeholder concerns.}
    \label{fig:useCases}
\end{figure}

We started the application of our method with the modeling activities within the problem space. First, we scanned the 67 stakeholder requirements that were provided by one OEM to identify use-cases for the system in development. 
This activity was supported by the system view \emph{use cases and application scenarios} that was implemented as a tree structure diagram as shown in Fig.~\ref{fig:useCases}. 

As a result, we identified the stakeholder concern: \emph{as a user I want to configure the SOC that shall be reached at a certain time}, 
which we assigned to the use-case: \emph{the user configures SOC and departure time on an HMI.} Based on the information given from the 67 stakeholder requirements, we identified 8 application scenarios which we linked to the use case. 
In addition to the scenario that describes a user interaction with the HMI, the identified scenarios also consider how to handle implausible data and errors in the power supply of the ECU. A more comprehensive scenario describes how implausible data leads to a deactivation of the timer-charging functionality that can be reactivated under specific constraints (self-healing). 
As a result of the activities \emph{model stakeholder and use-cases} and \emph{model stakeholder requirements}, we were able to link all 67 stakeholder requirements to at least one application scenario. 

\begin{figure}[ht]
    \centering
    \includegraphics[width=1.0\linewidth]{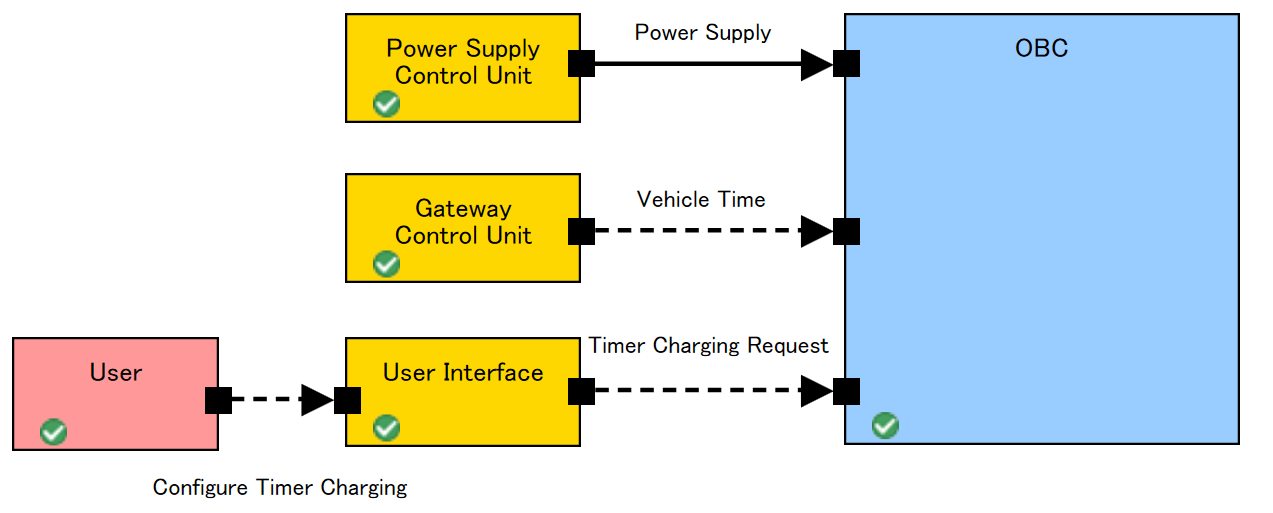}
    \caption{Context diagram including the user, interacting systems, and the system in development.}
    \label{fig:contextdiagram}
\end{figure}

Based on the identified use-case, we executed the activity \emph{model system context} to identify and model all interacting systems that are required for the identified use-case and the connected application scenarios.
We identified the \emph{OBC} as the system in development, and a \emph{user interface}, a \emph{gateway control unit}, and a \emph{power supply control unit} as interacting systems, which we added to the system model using a context diagram in iQUAVIS as shown in Fig.~\ref{fig:contextdiagram}. 

\subsubsection{Validation System Modeling}
\begin{figure*}[b]
    \centering
    \includegraphics[width=1.0\linewidth]{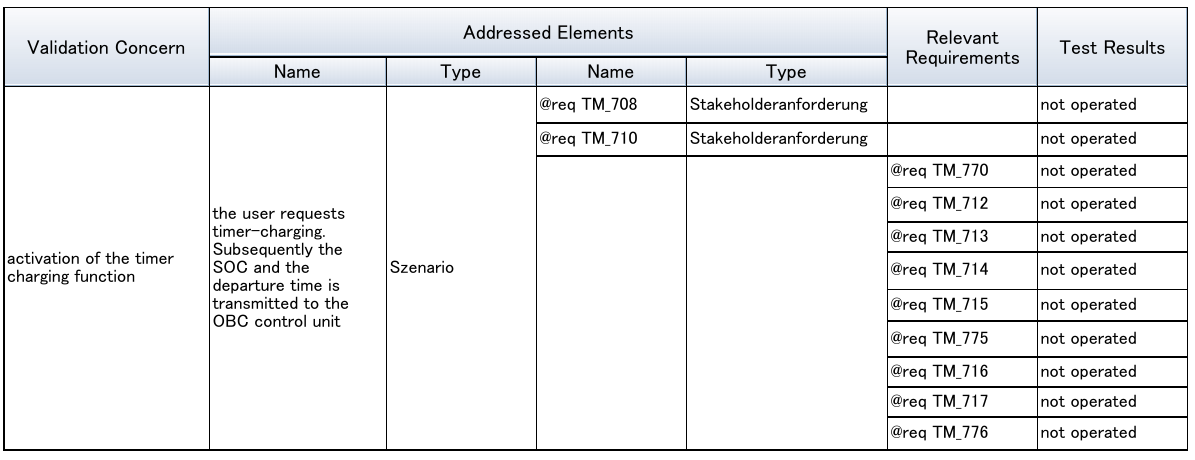}
    \caption{Validation concerns overview including addressed and relevant system elements and test results from previous iterations.}
    \label{fig:validationConcerns}
\end{figure*}
As an entry point for the validation system modeling, we used the resulting tree structure diagrams from the problem space modeling to define validation concerns and create links between these validation concerns and elements within the problem space. As a result of the activity \emph{derive validation concerns}, we were able to derive the view shown in Fig. \ref{fig:validationConcerns} from our system model. 
This view automatically lists all identified validation concerns (e.g., \emph{activation of the timer charging function}) from the model and shows elements that are related, i.e. linked in the model, to this concern. Specifically, we defined that the validation concerns can directly be rooted to one application scenario and two stakeholder requirements. In addition, the view shows that more stakeholder requirements are related to the addressed application scenario but not directly linked to the validation concern. Consequently, these requirements were shown as potentially relevant. In addition to the addressed elements and potential relevant elements, the view also includes test results from previous iterations, that can be considered when defining new validation concerns.  

With the information summarized in the validation concerns overview (Fig. \ref{fig:validationConcerns}) we entered the activity \emph{consolidate validation goals and tests} to arrive at concrete test cases and to define how these test cases contribute to specific validation goals. By applying the test case generation technique introduced in previous work\cite{Wiecher2021c,Fischbach2022}, we generated 111 test cases based on the 57 functional requirements and added these test cases to our system model by using a tree structure diagram. The aim of this activity was to ensure that each stakeholder requirement is addressed by the right amount of test cases. Consequently, we took each requirement and generated the test cases without focusing on dependencies between requirements. Subsequently, we combined the available test cases to define test scenarios that considered causal dependencies between single test cases and in this way addressed specific validation goals.    

\begin{figure*}[b]
    \centering
    \includegraphics[width=1.0\linewidth]{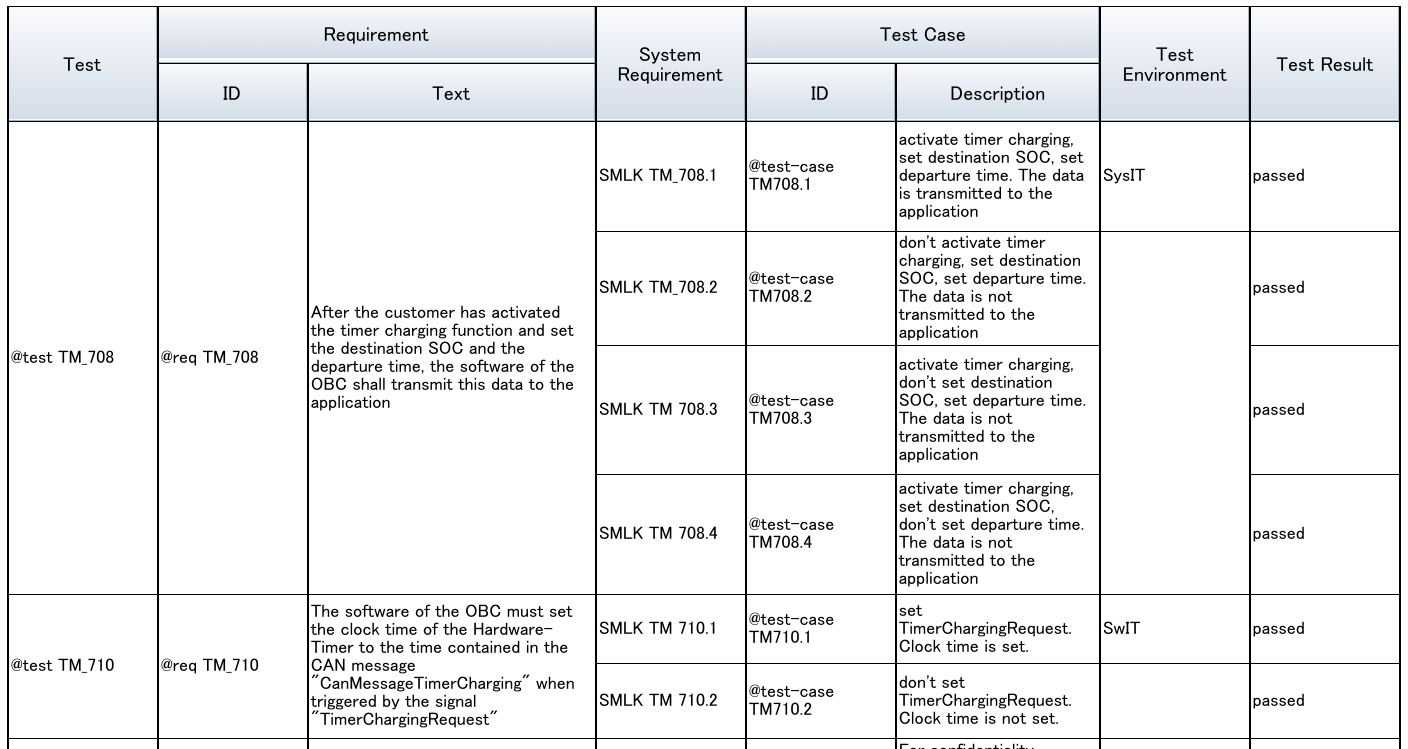}
    \caption{Overview derived from the system model containing all test cases for each stakeholder requirement with the related test environments and test results.}
    \label{fig:testCaseSpecification}
\end{figure*}
In addition to the definition of tests and validation goals, we explicitly modeled on which environments the tests should be executed by entering the activity \emph{model test environments}.  
At the case company, the timer-charging function was validated on two test environments that were operated in two different
departments. The first test environment was used for software integration testing and the second environment was used for system integration testing (cf. to the ASPICE processes SWE.5 and SYS.4 - see background information in Sect. \ref{sect:testspecification}).
The test environments are in turn composed of \emph{test environment objects} (see Fig. \ref{fig:ontology}), which we used to model specific resources (e.g., debug- or CAN interfaces). With this information we were able to link the different test environments to the already defined tests. Specifically, one test case required to read back the clock time that was written to the hardware-timer, which is only possible with access to internal ECU data. Therefore, we connected the corresponding test to the test environment for software integration testing, which included the test environment object \emph{debug-interface}, necessary to read the internal data. 

With executing the activity \emph{consolidate validation goals and tests} and \emph{model test environments}, we created the foundation for the activities \emph{requirements verification} and \emph{requirements validation}. These activities were based on two separate views, which we derived from our system model automatically.
We used both views for the test-driven modeling of system requirements. As a first step we used the view shown in Fig. \ref{fig:testCaseSpecification} to model each test case and run the TDSS process (see Sect. \ref{sect:smlk}). 

In this way, we were able to model all functional requirements driven by test cases, where the overview in Fig. \ref{fig:testCaseSpecification} includes the stakeholder requirements (e.g., @req TM\_710) which we used for the test case generation (e.g., @test case TM710.1), and the related system requirement as a result of our test-driven modeling (e.g., SMLK TM 710.1). Together with the linked test results we were able to document that the current state of the system requirements specification covers each stakeholder requirement and that each stakeholder requirement is verified.

\begin{figure*}[t]
    \centering
    \includegraphics[width=1.0\linewidth]{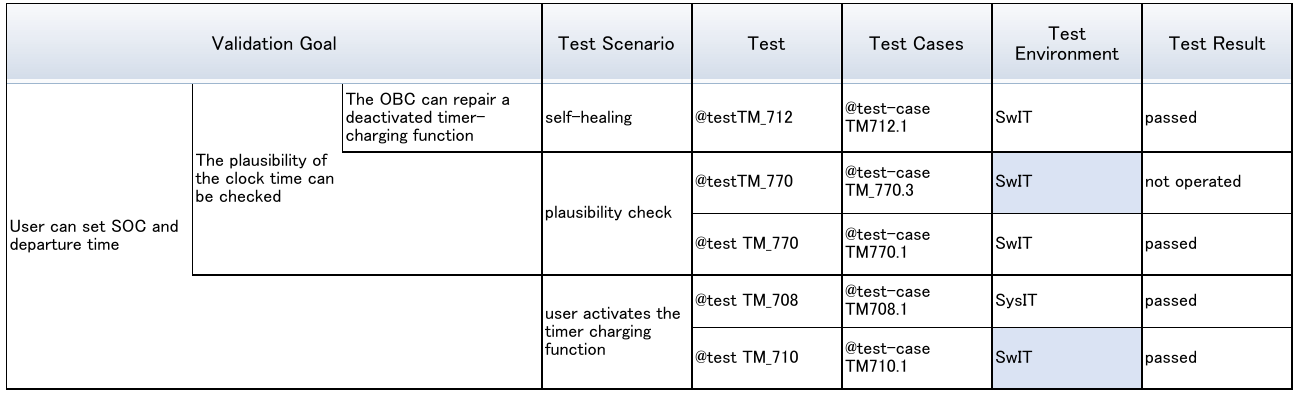}
    \caption{Validation goals that are derived from initially defined list of validation concerns. With the help of test scenarios, validation goals are linked to concrete test cases.}
    \label{fig:testGoalSpecification}
\end{figure*}

To additionally answer the question if the previously created specification was suitable to fulfill the validation goals, we executed the activity \emph{requirements validation} based on the view shown in Fig. \ref{fig:testGoalSpecification}. According to the list of all identified validation concerns, this view specified concrete validation goals and their dependencies. E.g., for the first goal it was requested that the user can set the SOC and the departure time. To reach this goal we defined one test scenario composed of two test cases, that we already modeled in the requirements verification activity. 
Depending on the first goal, we specified that the plausibility of the clock time can be checked as a second goal. And a third goal specified that if the plausibility check fails, a self-healing functionality can repair the deactivated timer-charging function. Consequently, with the resulting structure as shown in Fig. \ref{fig:testGoalSpecification}, we were able to identify which test cases contribute to the respective validation goal. 
For the requirements validation, we triggered the previously created scenario specification with the test scenarios that were linked to the consolidated validation goals. In this way we were able to demonstrate if the implemented systems requirements are sufficient to fulfill the validation goals \footnote{For details on the applied requirements modeling language within the requirements verification and validation activity, please refer to Appendix \ref{appendix:SMLK}}. 


\subsection{Expert Feedback}
To conduct the expert feedback, we used the previously described walkthrough and elaborated the outcome with two senior managers that were involved in the development of the timer-charging function following the traditional development process. The expert feedback was structured with the help of a semi-structured interview. This interview was based on related work that investigates the acceptance of MBSE approaches in general. The different categories of the questionnaire are based on Mandel et al. \cite{Mandel2022}. 
Basically, we divided the questions into two categories (cf. \cite{lohmeyer2014}). The first category deals with individual acceptance of our approach, and the second category includes topics that indicate organizational acceptance. 
For both the individual and organizational acceptance, our questions are based on sub-categories that were identified by related work and can be seen as fields of action to investigate the acceptance when developing MBSE methods \cite{Mandel2022}. Table \ref{tab:survey} provides an overview of all sub-categories and the related work from which these categories are derived.

\begin{table}
\begin{tabular}{m{4cm}| m{2cm} }
\textbf{Category - individual acceptance} & \textbf{derived from} \\
\hline
Perceived performance of individual users & \cite{Dumitrescu2021} \cite{bretz2019} \cite{gausemeier2015} \cite{lohmeyer2014} \cite{Matthiesen2014}   \\
\hline
Intuitiveness of applicability & \cite{lohmeyer2014}  \\
\hline
Flexibility and adaptability of the methodology & \cite{chami2018} \cite{Dumitrescu2021} \cite{friedenthal2009} \cite{lohmeyer2014} \cite{gausemeier2015} \\
\hline
Usability of the modeling tool & \cite{albers2013} \cite{chami2018} \cite{Dumitrescu2021} \cite{gausemeier2015} \cite{lohmeyer2014} \cite{Matthiesen2014}\\
\hline
Target vision and modeling procedure clear for users &  \cite{Dumitrescu2021} \cite{chami2018} \cite{friedenthal2009}\\
\hline
Appropriate level of formalization and ontology to support (and not hinder) communication among users &  \cite{gausemeier2015} \cite{tschirner2015} \cite{albers2013} \\
\hline
\textbf{Category - organizational acceptance} & \textbf{derived from} \\
\hline
(Monetary) benefit-cost/effort ratio of applying the methodology &  \cite{bretz2019} \cite{chami2018} \cite{Dumitrescu2021} \cite{gausemeier2015} \cite{lohmeyer2014}\\
\hline
Teach- and Learnability of the methodology & \cite{gausemeier2015} \cite{lohmeyer2014} \cite{friedenthal2014} \cite{bone2010} \\
\hline
Reusability and extendibility of the method and created models & \cite{lohmeyer2014} \cite{chami2018} \\
\hline
Problem orientation/support in modeling the problem space &  \cite{Cloutier2019} \cite{tschirner2015}\\ 
\vspace{0.1cm}
\end{tabular}
\caption{\label{tab:survey}Identified categories to structure the online-survey and the expert interviews, adapted from Mandel et al. \cite{Mandel2022}.}
\end{table}

\subsection{Student Projects \& Online Survey}
In addition to the application at our industry partner, we applied our technique within three student projects. As part of the masters program \emph{embedded systems engineering} at \emph{dortmund university of applied sciences and arts}, 11 students worked on 3 projects for 1 week. These projects were coordinated with 2 industry partners (Kostal, Two-Pillars). 
At the beginning of the week, the students defined the individual projects and the project goals together with Kostal. In addition, TwoPillars provided an introduction to the modeling tool iQUAVIS. The modeling approach was introduced with the help of the example introduced in Sect. \ref{sect:example} and the previously described walk-through. The complete week was structured in 5 milestones with presentations and discussions at the end of each day. At the end of day 1, the students presented their individual project outlines, the focus of day 2 was on problem space modeling (I1 in Fig. \ref{fig:mbsescil}), and day 3 and 4 focused on the test-driven requirements modeling and analysis (I2 and I3 in Fig. \ref{fig:mbsescil}). The last day was used for the final presentations and to conduct the online survey.

\subsection{Survey Results}
To capture the feedback from the expert interviews and from the student projects, we provided statements for each identified category, where the  respondents could make a choice in the range of 1 (do not agree) and 4 (fully agree). The results are summarized in Table \ref{tab:surveyresults}. The feedback for each category is discussed subsequently. 

\begin{table*}[]
    \centering

\begin{tabular}{ |p{3cm}|p{6.2cm}||p{1.2cm}|p{1.2cm}|p{1.2cm}|p{1.2cm}| }

\hline
\textbf{Identified category} & \textbf{Derived statement} & \textbf{1 (do not agree )}& \textbf{2} & \textbf{3} & \textbf{4 (fully agree)} \\
\hline
\multirow{2}{3cm}{\textbf{Perceived performance of individual users}}  & I recognize an added value of MBSE for my project     & 0 &   0 &   5 (4s, 1e) &   8 (7s, 1e)\\
\cline{2-6}
                                                    & The provided methodology helps me to navigate to the exact views/activities where MBSE can be helpful in my project     & 0 &   0 &   4 (2s, 2e) &   9 (9s)\\
\hline
\textbf{Intuitiveness of applicability}  & The developed training concept and the application of the framework help to find a quick start into modeling     & 0 &   0 &   4 (4s) &   7 (7s)\\
\cline{2-6}
                                                    & With the methodology used, I can quickly and specifically find the modeling activities and diagrams that are relevant to my modeling purpose/problem     & 0 &   0 &   6 (4s, 2e)&   7 (7s)\\

\hline
\textbf{Flexibility and adaptabi\-lity of the methodology}  & The modeling activities can be performed flexibly/iteratively without following a strict waterfall process, thus supporting a goal-oriented modeling      & 0 &   0 &   5 (3s, 2e) &   8 (8s)\\

\hline
\textbf{Usability of the modeling tool}  & By using the provided template and the depictions of the framework, the navigation in the software tool is made easier     & 0 &   0 &   4 (3s, 1e) &   8 (8s)\\
\cline{2-6}
                                                    & In the provided template and software tool, all functions needed for modelling in my project are explained and can be used without extensive training     & 0 &   3 (3s) &   7 (6s, 1e) &   2 (2s)\\

\hline
\multirow{2}{3cm}{\textbf{Appropriate level of formalization and onto\-logy to support (and not hinder) communication among users}}  & The model can be used for communication in the team      & 0 &   0 &   6 (5s, 1e) &   7 (6s, 1e)\\
\cline{2-6}
                                                    & The model can be used to communicate with external development partners (supervisor at milestones)    & 0 &   1 (1s) &   4 (3s, 1e) &   8 (7s, 1e)\\
\cline{2-6}
                                                    & Modeling with given elements (from the ontology) supports unambiguous modeling and unambiguous communication in the team    & 0 &   0 &   5 (5s) &   6 (6s)\\
\cline{2-6}
                                                    & The applied scenario-based modeling has fostered coordination on expected system behavior within the team    & 0 &   0 &   6 (4s, 2e)&   7 (7s)\\
\hline
\textbf{Problem orientation and modeling depth}   & The test-driven modeling and the elements of the validation system (e.g., validation goals) were helpful to decide what needs to be modeled and to which extend      & 0 &   0 &   3 (2s, 1e) &   10 (9s, 1e)\\
\cline{2-6}
                                                    & When following the test-driven modeling, the resulting sequence diagrams specified our systems requirements sufficiently and addressed the identified problem    & 0 &   0 &   2 (2s) &   10 (9s, 1e)\\
\hline
\textbf{Benefit-cost/effort ratio of applying the methodo\-logy}   & By targeted guidance to modeling activities and viewpoints through the methodology, modeling effort can be kept reasonable in relation to the overall project      & 0 &   1 (1s) &   5 (3s, 2e) &   7 (7s)\\
\hline
\textbf{Teach- and Learnability of the methodology}   & The structure of the training is well comprehensible      & 0 &   1 (1s) &   7 (6s, 1e) &   4 (4s)\\
\hline
\multirow{2}{3cm}{\textbf{Reusability and extendibility of the method and created models}}   & The application of the method is not limited to a specific project, it can be applied to different projects     & 0 &   0 &   6 (4s, 2e) &   7 (7s)\\
\cline{2-6}
                                                    & The structure of the methodology (problem space, validation system, product) supports the reuse of the created model elements in subsequent development projects    & 0 &   0 &   5 (4s, 1e) &   7 (7s)\\
\hline
\end{tabular}
\caption{\label{tab:surveyresults}Results from the online survey. For each statement, the table shows the number of responses for a choice from 1 (do not agree) to 4 (fully agree). 
In the parentheses, a distinction is made between the responses of the students (s) and the experts (e).}
\end{table*}

\textbf{\emph{Perceived performance of individual users}}:

Both the students and the experts reported that they see an added value of our methodology for their projects. However, based on the previously described walk-through, the experts stated that the added value depends on the project size. 
Although the central entry point (Fig. \ref{fig:v&vframework}) and the identified views and activities were seen as supportive to guide the modeling tasks, for small projects it may be to much effort. Nevertheless if multiple validation systems and complex functionalities are involved, they fully agree that the application of the methodology is beneficial.

The students stated that the structuring of the activities and the associated views were helpful to directly start modeling a current problem while keeping the overall model in mind.

\textbf{\emph{Intuitiveness of applicability}}:

The students agreed that with the provided training (introduction by the tool vendor - Two Pillars, and an introduction of the methodology based on an example from the industry partner - Kostal, combined with discussions at defined milestones) they found a quick start into the modeling by addressing their individual technical problems. 

Since the two experts were not directly involved in the student projects and their opinion is based on experiences from the walk-through, they could say little about the intuitiveness of applicability. However, they agreed that with the methodology it is easier to find the required modeling activity and diagram, compared with existing approaches at the case company. 

\textbf{\emph{Flexibility and adaptability of the methodology}}:

Regarding the flexibility and adaptability, the students and experts agreed that the the central entry point within the modeling tool (Fig. \ref{fig:v&vframework}) supported an iterative development and can be applied specifically for the current problem within the project.
The students reported that they started modeling within the problem space based on their initial project outline, and, after continuing on day two and three with the validation system and product modeling, they often jumped back into the problem space to discuss use cases, scenarios and related stakeholder requirements. 

One of the experts mentioned, that the problem space modeling is usually done by the OEM, and the missing information is often exchanged informally based on requests of the supplier. With explicitly capturing this information with the help of context diagrams, use cases and scenarios, they agreed that this is beneficial to support a goal oriented modeling, especially for the validation system modeling including the definition of validation goals and test scenarios.

\textbf{\emph{Usability of the modeling tool}}:

The usability was considered as positive. The students stated that the structuring of activities helped to find what modeling task should be executed.
Since the experts did not use the modeling tool extensively, they could say little about the usability. 
Regarding the provided template and the different functionalities of the modeling tool some students reported that more training would be necessary. One expert mentioned that structuring activities within the tool, as depicted in Fig. \ref{fig:v&vframework}, can enhance applicability. Furthermore, it was confirmed that the available block diagrams, tree structures, and textual descriptions can be easily utilized. 

\textbf{\emph{Appropriate level of formalization and ontology to support (and not hinder) communication among users}}:

Overall, students and experts agreed that the degree of formalization and ontology supported the communication within the three project teams, or can support the communication within engineering teams respectively. 
Also for the communication with external development partners, the majority of respondents agreed that the resulting system model can be used for communication. 
Within the student projects, the students quickly adapted the methodology and used the ontology elements actively to discuss the current modeling problem. Especially the validation system modeling and the concept of test-driven system requirements modeling was extensively applied with the results being used for discussions within the teams. Consequently, they answered the statement positively. 

However, the experts could not say if the elements of the ontology would actively be used within engineering teams for the communication within and across teams. They stated that the application needs to be investigated in several and larger projects.

For the scenario-based modeling, students and experts agreed that this way of modeling within the problem space and for the modeling of system requirements fostered or can foster the coordination on expected system behavior within the teams. 

\textbf{\emph{Benefit-cost/effort ratio of applying the methodology}}: 

The majority responded that the effort can be kept reasonable in relation to the overall project. However, according to the experts, this strongly depends on the project size as already mentioned above. For large projects, the ability to derive specification as shown in Fig. \ref{fig:testCaseSpecification} and Fig. \ref{fig:testGoalSpecification} is seen as beneficial, where for small projects the modeling effort is seen as to high.

\textbf{\emph{Problem orientation and modeling depth}}: 

The students and one expert fully agreed that the test-driven approach with the early definition of validation concerns and validation goals, leading to a test-driven development of system requirements, were helpful to decide what needs to be modeled and to which extend. 
The statement regarding the application of our model-based requirements specification technique goes in the same direction. Most of the respondents fully agreed that the test-driven modeling led to clearly defined systems requirements that addressed the identified problem.

\subsection{Discussion}

The connectivity of current automotive systems leads to often changing stakeholder requirements\cite{Schnitzler2020}. For practitioners it is challenging to systematically analyze and specify system requirements when the environment and interconnected systems change during development time (cf. Sect. \ref{sect:example}), which leads to changing stakeholder request and consequently to an iterative system development \cite{Kasauli2021}. Given this practical challenge, this work addresses how complete and consistent requirement and test specifications can be created, when considering an iterative development and the demands of automotive development standards (a.o. traceablity and verifiability of requirements in a plan-driven and stage-gate oriented development process, cf. Sect. \ref{sect:requirementspecification} and Sect. \ref{sect:testspecification}). 

Based on the findings that challenges in requirements engineering and verification \& validation are predominantly investigated separately in both research and practice\cite{Bjarnason2013,Bjarnason2019}, this article is based on the idea of integratively analyzing and specifying requirements and tests.

Considering that many model-based attempts have been made in the field of requirements and test specification, but the adaption in industry is relatively low \cite{Liebel2019},  our contribution focuses on a application-oriented technique.

According to existing MBSE approaches (cf. Sect. \ref{sect:Model-Based Systems Engineering (MBSE)}), our technique (Sect. \ref{sect:systemsmodeling}) is based on a central system model combined with scenario-based modeling techniques and suitable automation approaches (cf. Sect. \ref{sect:smlk}). 

To investigate the feasibility of our approach, we applied our technique at our industry partner and within student projects. Besides the detailed feedback summarized in Table \ref{tab:surveyresults}, one key insight is that despite the overall complexity of the method and the modeling language used, the students were successfully able to realize and present their individual projects within only one week of work and without expert knowledge. One reason for this may be the implementation of the central entry point for all modeling activities (see Fig. \ref{fig:v&vframework}), which provides orientation, and combined with the connected system views covers the complexity of dependencies within the system model (see categories \emph{perceived performance of individual users}, \emph{intuitiveness of applicability} and \emph{appropriate level of formalization and ontology to support communication among users} in Table \ref{tab:surveyresults}).  

From an industry perspective the responses for the category \emph{problem orientation and modeling depth} provides valuable insights. When modeling complex systems, it is often challenging to decide what to model and to which extend. Since our technique follows a test-driven approach combined with the explicit definition of validation concerns and validation goals, we guide the modeling activities and ensure that the modeled system requirements are traceable to a validation and hence stakeholder concern. Although it is possible to model the system behavior in-depth using the applied modeling language, the iterative modeling method (TDSS) combined with the related system views (Fig. \ref{fig:testCaseSpecification} and Fig. \ref{fig:testGoalSpecification}) supports the systems engineer in iteratively deciding what needs to be modeled next. The related statements to the category \emph{problem orientation and modeling depth} were answered very positive. 

Despite the overall positive feedback, there is a continued need for application, validation, and adaptation of the approach in complex development projects. While the implemented example is based on real requirements of a complex function, the resources and time available within a research project are not sufficient to address all the details that arise in real development projects. With the artifacts provided in this\footnote{
\url{https://mbse4u.com/sysml-tools/}\\
\url{https://www.two-pillars.de/iquavis/}\\
\url{https://bitbucket.org/jgreenyer/workspace/projects/SMLK}\\
\url{https://bitbucket.org/jgreenyer/smlk-animator}\\
\url{http://www.cira.bth.se/demo}} and previous work \cite{Wiecher2019, Wiecher2021, Mandel2021, Fischbach2022}, we encourage others to implement, adapt and further evaluate the presented technique for different development situations.

\section{Closing and Next Steps}
\label{sect:closing}

Following the DSR paradigm, we integrated different techniques and concepts and designed and applied a new DSR artifact at a Tier1 supplier company to address a practical problem: the specification of system requirements and tests for complex automotive systems. It is obvious that the individual techniques are studied comprehensively and in depth in individual research directions (cf. scenario-based modeling \cite{Greenyer2021} or validation-focused MBSE \cite{Mandel2021}). However, none of these works alone is able to solve the identified practical problem. The contribution of this work is the integration of the single concepts to arrive at complete and consistent requirements and test specifications in an iterative development context, which is driven by changing stakeholder requests. By the combination of recent research results as part of the MoSyS research project, we found the the new DSR artifact extends our previous work and finally can be applied beneficially in a real world context. The next steps focus on the roll-out of our technique in different development projects.

\bibliography{references}





\end{document}